\documentclass{revtex4}
\usepackage{bm}
\usepackage{amsmath}
\usepackage{epsfig}
\usepackage{graphicx}
\usepackage{pdfpages}
\usepackage{color}   

\begin{document}
	
	\title{ Gamma production in neutrino interaction with nuclei }
	\author{G. Chanfray, } 
	\affiliation{Univ Lyon, Univ Claude Bernard Lyon 1, CNRS/IN2P3, IP2I Lyon, UMR 5822, F-69622, Villeurbanne, France}
	\author{M. Ericson} 
	\affiliation{Univ Lyon, Univ Claude Bernard Lyon 1, CNRS/IN2P3, IP2I Lyon, UMR 5822, F-69622, Villeurbanne, France}
	\affiliation{Theory division, CERN, CH-12111 Geneva }
	
	\begin{abstract}
		
		We evaluate the cross-section for gamma production by neutrinos through a meson exchange effect which derives from the concept of axial-vector mixing. The resulting cross-section leads to some increase of the  gamma production cross-section by neutrinos, especially at low  neutrino energies, which may influence the understanding of the  low energy excess of electron-like events seen in the MiniBooNE experiment.
	\end{abstract}
	
	\maketitle
	\section{Introduction}\label{Intro}
	
	The problem of gamma emission in the interaction of neutrinos with nuclei is of a great interest for the interpretation of the low energy excess of electron-like events seen in the MiniBooNE experiment \cite{MiniBooNe2009,MiniBooNe2013}. In this short baseline (541 m) muon neutrino primary beam experiment a number of low energy electrons have been detected in the  target. The shortness of the baseline excluding oscillations into the known neutrinos, this observation has been interpreted as an evidence for a new neutrino, the sterile neutrino, more massive than the known neutrinos and whose existence has been vividly discussed. This anomalous excess has been confirmed in a recent analysis \cite{MiniBooNe2018}. The importance of this result has triggered a number of investigations to explore alternative interpretations for the presence of these electrons. A possibility is that gamma rings produced in the target have been mistaken for electron rings, as the MiniBooNE detector cannot distinguish the two, thus artificially increasing the apparent number of electrons. It is therefore crucial to have a proper evaluation of the gamma emission background in the interaction of neutrinos. \\
	Several evaluations of this process have been made \cite{Hill2010,Hill2011,Serot2012,Serot2013,Wang2014,Wang2015}. They involve in particular the production of a Delta, which decays by gamma emission. In these events where the gamma rings could be mistaken for electrons, no lepton are emitted  and the process for gamma production  involves a neutral current transition: the vector boson $Z_0$  in its interaction with a nucleon excites a Delta which emits a photon, as shown in Fig 1a. However, even if all these photons are mistaken for electrons this process has not been able to account for the observed number of electrons without a sterile neutrino oscillation \cite{Katori2020}.\\ 
	
	In this  work we introduce another source for gamma production by neutrinos which, to our knowledge, has not been considered in connection with this problem. It involves a meson exchange effect from a contact vertex where, together with the gamma, a pion is produced at the $N\Delta$ vertex as depicted on Fig. 1b. The contact $\langle N| \pi \gamma |\Delta\rangle$ coupling is obtained from the usual p wave coupling $\langle N|\partial_\nu\pi |\Delta\rangle$ by the minimal  substitution where in short  $\partial_\nu$ is replaced by $(\partial_\nu - q\,A_\nu)$. Its precise form will be given below.
	In the interaction of real photons with a free nucleon this contact coupling is responsible for an appreciable part of the photon-nucleon cross-section in the energy region above the Delta energy,  
	$\omega\simeq 400 - 500\ MeV$. Notice that the process of $\gamma \pi$ simultaneous production as such would not affect the MiniBooNE interpretation as the pion produced in this process is detectable, while instead in the selected events no pions are observed. However in the nucleus the pion is dressed, in particular by particle-hole ($ph$) excitations, thus acquiring a broad spectrum which extends on the low energy side of the pion mass. In the MiniBooNE detector which does not identify nuclear excitations, when a pion produced is disguised as a $ph$ state it becomes invisible, simulating a simple gamma production process and contributing to an increase of the gamma emission  cross section. The fact that  the energy distribution of the process where the pion is materialized as a $ph$ excitation is smaller than  the one where it is emitted as a real pion is interesting, as the apparent electron excess in MiniBooNe occurs below a reconstructed neutrino energy of approximately $400 \,MeV$.\\
	The introduction of the graph of Fig. 1b is not randomly chosen: it embodies the concept of  axial-vector mixing in the nuclear medium introduced by Chanfray et al \cite{CDER99}. This concept consists in the following: the s wave absorption (or emission) by a nucleon (or a Delta) of a pion emitted by a neighboring nucleon, as  illustrated in Fig 1b, produces a parity change which transforms a vector correlator into an axial one (or the reverse). It extends to the nuclear case the  concept, introduced by Dey et al \cite{DEI90}, of parity mixing by  the thermal  pions of a heat bath. In the nuclear medium the virtual pions of the pionic clouds replace the thermal pions of the heat bath. Drawn in the perspective of correlator mixing the vector-vector correlator (graph 2c  of Fig. 2  of ref \cite{CDER99}) is associated with the mixing ones represented by graphs 2d-e-f of the same figure. As a side remark we also point out that the pion cloud contribution to the broadening of the rho meson observed in dilepton production in relativistic heavy ion collision \cite{CS93,CRW96} originates from such an axial-vector mixing effect, as demonstrated first in this reference \cite{CDER99}.\\
	In the simpler case of the interaction of real photons with nuclei the effect expected from the correlator mixing is a spreading of the gamma nucleus cross section due to the $\pi N\Delta$ contact term over a larger range of energy, increasing  the nuclear cross section as compared to the nucleonic one in the low energy transfer region, $\omega\leq 500\, MeV$ and producing some depletion above, in the energy region  where the original $\pi N\Delta $ contact  term acts. These features are remarkably similar to those of the MiniBooNE excess event but this analogy is premature, as the kinematics of neutrino interactions is  different.
	The contact $\pi N\Delta$ graph that we have discussed is not the only mixing term. There is also the one where the  gamma is absorbed by a pion in flight as shown in Fig. 1c. We have also  evaluated  this term and it contributes significantly. \\
	
	Summarizing, the aim of the present work is the evaluation of the cross section for the process of gamma emission on the $^{12}C$  nucleus  by  neutrinos  through the contact $N \Delta \gamma\pi$ term, leading to a final $2p-2h$ excited state. 
	We use for the $^{12}C$ nucleus a nuclear matter description with a typical density, $\rho=0.8 \,\rho_0$, where $\rho_0$ is the normal nuclear matter density.  
	The process which  for a free nucleon is :  
	$Z_0 + N \,\rightarrow\,\Delta \,\rightarrow\, N\pi\gamma$
	leads in the nucleus to  :
	$Z_0 + A\,\rightarrow\,\Delta \,(A-1) \,\rightarrow\,\gamma\pi \,1p1h\,(A-1) \,\rightarrow\,\gamma \,2p2h\,(A-2)$.
	As mentioned above, the  final state which consists of a gamma plus an (invisible)  $2p-2h$ excitation can be mistaken for a  gamma emission and hence for an electron emission process. This contribution is appreciably different in different neutrino energy regions, an interesting feature for the MiniBooNE experiment where the excess is concentrated in a region of neutrino energy below $\simeq 400 MeV.$ \\ 
	
	\begin{figure}
		\centering
		\includegraphics[width=0.8\textwidth]{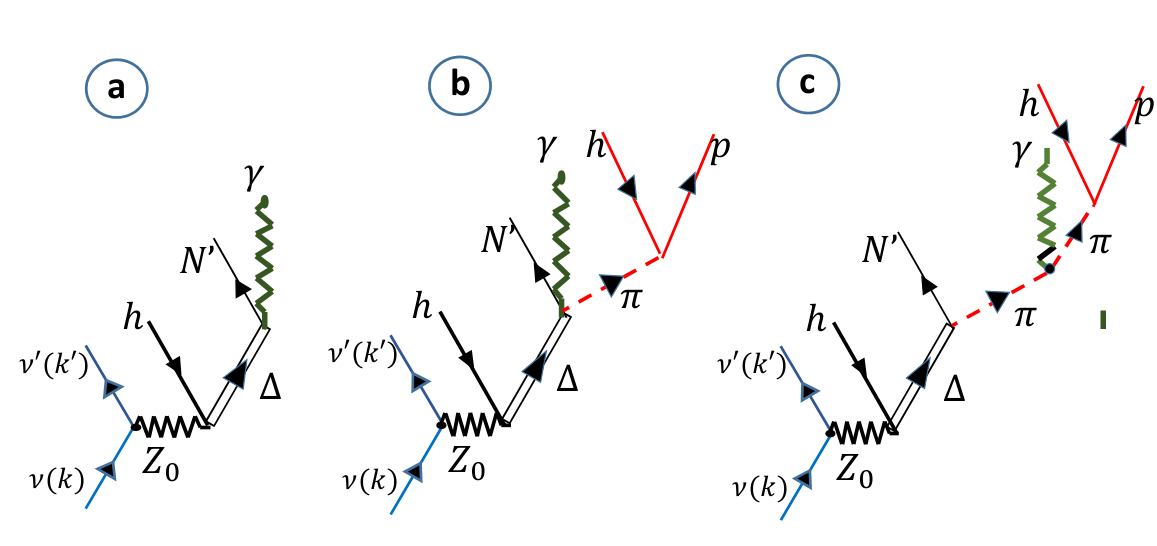}
		\caption{Left panel (a): single $\gamma$ emission process considered in \cite{Hill2010,Hill2011,Serot2012,Serot2013,Wang2014,Wang2015}. Middle panel (b): contact $\gamma$-pion emission process. Right panel (c): Pion in flight process.}
		\label{f1}
	\end{figure}

	In the following we denote by $ k=(E_\nu, {\bf k}) $ and $k'=(E'_\nu, {\bf k}')$ the quadrimomenta of the incoming and outgoing neutrinos, $\omega=E_\nu-E'_\nu$ and   ${\bf q}={\bf k}-{\bf k}'$, the energy and  momentum transferred to the nucleus and $\theta$ the scattering angle. The  $Z_0 N\Delta$ coupling  can be extracted from the isovector piece of the spatial part of  the $\bar q q Z_0$ vertex ; with standard notation it reads:
	\begin{equation}
		{\cal L}_{\bar{q} q Z_0}=\frac{g_2}{2\cos \theta_W}\,\left(V_\mu(x)-A_\mu(x)\right)\,Z_0^\mu(x)\simeq \frac{g_2}{2\cos \theta_W}\bar q(x)\,\gamma_j\,\left(1 -\gamma_5\right)\, t_3^q \,q(x)\, Z_0^j(x) .
	\end{equation}
	At low neutrino energy the dominant part is the hadronic axial current, i.e., the $\gamma_5$ piece. The vector current piece  gives a relativistic correction which generates  an axial-vector interference in the cross-section resulting in a higher cross-section for neutrinos than for antineutrinos. For the single gamma production discussed below this interference gives typically a $30\%$ effect compared to the averaged cross- section where only the axial piece is taken into account. This point is discussed in some details in section IV of Ref. \cite{Hill2010} and is visible on Fig. 4 of Ref.  \cite{Hill2010}, on Fig. 3 of Ref. \cite{Serot2013} or on Fig. 5 and 8 of Ref. \cite{Wang2014}.  In this exploratory work devoted to the comparison of single gamma process with the pion exchange process we  limit ourselves to the axial coupling such that, strictly speaking, our result applies to the average sum of the neutrino and antineutrino cross-sections. Hence in the following we take:  
	\begin{equation}
		{\cal L}_{\bar{q} q Z_0}\simeq -\frac{g_2}{2\cos \theta_W}\bar q(x)\,\gamma_j\,\gamma_5\, t_3^q \,q(x)\, Z_0^j(x) .
	\end{equation}\\

	The cross-section for gamma production induced by a neutral current which excites a Delta from a nucleon with momentum $p$   producing  a photon in the final state writes:
	\begin{equation}
		d\sigma=2 G^2_F\frac{2\pi\, dcos\theta\,k'\,E'_\nu\,dE'_\nu}{(2\pi)^3\,2E_\nu\,2E'_\nu}\,L_{\mu\eta}\,H^\mu H^{*\eta}\,2\pi\,\delta\left(E_{f\gamma }+E'_\nu-E_p-E_\nu\right),
		\label{sigmagen}
	\end{equation}
	with 
	\begin{equation}
		L_{\mu\eta}=Tr\left\{\not k\gamma_\mu(1-\gamma_5)\not k'\gamma_\eta(1-\gamma_5)\right\}=8\,\left(k_\mu k'_\eta +
		k_\eta k'_\mu -g_{\mu\eta} k\cdot k' \pm \,i\varepsilon_{\mu\eta\alpha\beta} k^\alpha k'^\beta\right)
	\end{equation}
	\begin{equation}
		H^\mu=\langle f\gamma |\int d{\bf x}\,\,{\bf j}_{em}({\bf x})\cdot{\bf A}({\bf x})\, |\Delta\rangle \,\,G_\Delta (\omega,{\bf p}+{\bf q})\,\,
		\langle \Delta |\int d{\bf y}\,\bar q({\bf y})\,\gamma^\mu\,\gamma_5\, t_3^q \,q({\bf y})\,e^{i{\bf q}\cdot{\bf y}}|p \rangle .
		\label{current}
	\end{equation}
	The first form of the leptonic tensor applies to the neutrino case whereas the second form,  with the explicit incorporation of the antisymmetric piece and where the plus (minus) sign refers to the neutrino (antineutrino),
	covers both $\nu $ and $\bar\nu $ cases. This  antisymmetric  piece, when contracted with the hadronic tensor, gives the axial-vector interference contribution to the cross-section that we have ignored. For what concerns the hadronic piece $H^\mu$, a summation over spin and isospin states of the intermediate delta is understood. For the  $\Delta$ propagator, neglecting Fermi motion, we take the simplified form:
	\begin{eqnarray}
		G_\Delta (\omega,{\bf q})&\simeq& \left[\omega-\sqrt{M_\Delta^2+q^2}\,+M_N\,+i\,\frac{\Gamma(\omega)}{2}\right]^{-1} \nonumber\\  
		\Gamma(\omega)&=&\frac{1}{6\pi}\left(\frac{g_A}{2 f_\pi} R_{N\Delta}\right)^2
		\left(\omega^2-m^2_\pi\right)^{3/2}\,\, \hbox{with} \,\,\; R_{N\Delta}=\frac{g_{\pi N\Delta}}{g_{\pi NN}}. 
	\end{eqnarray}
	In the numerical evaluations we will take for the ratio of the Delta and nucleon coupling constants : $ R_{N\Delta}=2$. As mentioned above, in the non relativistic limit the spatial part of the axial current dominates. We take it in the standard form:
	\begin{equation}
		\langle \Delta |\int d{\bf y}\,\bar q({\bf y})\,\gamma^j\,\gamma_5\, t_3^q \,q({\bf y})\,e^{-i{\bf q}\cdot{\bf y}}|p \rangle=\frac{G_A(Q^2=kk'\,\left(\cos\theta-1\right))}{2}R_{N\Delta}\,\langle \Delta |\,S_j\,T_3 \,|p \rangle ,
		\label{axialmatrix}
	\end{equation}
	where $S_j$, ($T_k$) are the spin (isospin) transition operators between the spin-isospin 3/2 $\Delta$ state and the spin-isospin 1/2 nucleon state  with reduced matrix element $\left\langle \frac{3}{2}\left\|S,T\right\|\frac{1}{2}\right\rangle=2$. In the right-hand side of Eq. (\ref{axialmatrix}) the Delta and proton states refer to the spin-isospin quantum numbers only. The axial form factor is taken in the standard dipole form with $G_A(0)=g_A =1.26$ and a cutoff parameter $M_A=1.032\,GeV$.\\
	\section{Single photon  emission cross-sections  off nuclei}\label{single cross-section}
	
	In the case of a single gamma emission (i.e., $\Delta\rightarrow \gamma N$)  with momentum ${\bf p}_\gamma$ and polarisation $\vec\epsilon_\lambda$, the $\gamma N\Delta$ vertex, i.e., the electromagnetic matrix element, is obtained \cite{Oset82} by assuming the existence of a scaling law between the nucleon and Delta axial and magnetic matrix elements:
	\begin{eqnarray}
		\langle N'(p') ;\gamma({\bf p}_\gamma,\vec\epsilon_\lambda) |\int d{\bf x}\,{\bf j}_{em}({\bf x})\cdot{\bf A}({\bf x})\, |\Delta\rangle &=&
		\frac{e}{\sqrt{2p_\gamma V}}\,\langle p'|\,\int d{\bf x}\,\,e^{-i{\bf p}_\gamma\cdot{\bf x}}\,\bar q({\bf x})\,\vec\gamma\cdot\vec\epsilon\left(\frac{1}{6}+t_3^q\right)\,q({\bf x})\,|\Delta\rangle\\
		&=&-\frac{i e\,R_{N\Delta}}{\sqrt{2p_\gamma V}}\,\frac{\mu_p-\mu_n}{4 M_N}\,\langle p'|\left({\bf S}^\dagger\times{\bf p}_\gamma\right)\cdot\vec\epsilon\,\,T_3^\dagger\,|\Delta\rangle\,\delta_{{\bf p}_\gamma +{\bf p}'-{\bf q}-{\bf p}}.
		\label{elec}
	\end{eqnarray}
	In the evaluation of the cross-section, one has to perform a summation over the spin-isospin states  of the intermediate Delta and of the final emitted nucleon (see appendix A). Ignoring Fermi motion, the resulting cross section per nucleon for gamma production  writes:
	\begin{equation}
		\frac{d\sigma^\gamma}{dE'_\nu d\cos\theta}=\left[\frac{G^2_F}{4\pi}\frac{k'}{k}\,8\,kk'\,\left(3\,-\,\cos\theta\right)\right]\,\left[\frac{4}{9}\,R_{N\Delta}^2\,\left(\frac{G_A(Q^2)}{2}\right)^2 
		\right] \left[\frac{1}{2\pi}\,\Gamma^\gamma(\omega)\right].
		\label{sigmasingle}
	\end{equation}
	Neglecting nucleon recoil the radiative Delta width $ \Gamma^\gamma (\omega)$  is:
	\begin{eqnarray}
		\Gamma^\gamma (\omega) &=&\frac{4}{9}\,R_{N\Delta}^2\,\left(\frac{e}{2 M_N}\right)^2 \left(\frac{\mu_p-\mu_n}{2}\right)^2
		\int\frac{d{\bf p}_\gamma}{(2\pi)^3\,2p_\gamma}p^2_\gamma\,2\pi\,\delta(p_\gamma +\epsilon_{p'}-\omega -\epsilon_p)\nonumber\\
		&\simeq& \frac{4}{9}\,R_{N\Delta}^2\,\left(\frac{e}{2 M_N}\right)^2 \left(\frac{\mu_p-\mu_n}{2}\right)^2\,
		\frac{\omega^3}{2\pi}.
	\end{eqnarray}

	The result of our evaluation for the single gamma emission off $^{12}C$ as a function of the neutrino energy is displayed on Fig. 2. We find a qualitative agreement with Ref. \cite{Hill2010,Hill2011,Serot2012,Serot2013,Wang2014,Wang2015}, certainly not perfect but  sufficient for the purpose of this article. For instance, for a neutrino energy of $E_\nu=\,0.3\, GeV$, we find for the cross-section per nucleon 
	$\sigma^\gamma/A\simeq\,  0.27\, 10^{-42} cm^2$ which is twice larger than the Delta contribution shown on Fig. 4 of Ref. \cite{Hill2010}, but  the cut of $200\,MeV$ applied to the photon energy in this reference could explain this difference. For a neutrino energy of $E_\nu=\,0.5\, GeV$, we find $\sigma^\gamma/A\simeq\,  2.\, 10^{-42} cm^2$ which is close, although slightly larger,  to the averaged $\nu \bar\nu$ cross-section shown  on the same figure (Fig. 4 of Ref. \cite{Hill2010}) and on Fig. 8 of Ref. \cite{Wang2014}. Here it is apparent that this Delta contribution  dominates over other processes such as  "Compton-like" scattering and omega exchange.
	At a higher neutrino energy,  $E_\nu=\,1\, GeV$, we find $\sigma^\gamma/A \simeq  8\,10^{-42} cm^2$,    again  slightly larger than  the averaged $\nu \bar\nu$ cross-section obtained in Refs.  \cite{Hill2010,Serot2013,Wang2014} where it is apparent that  the Delta contribution is largely dominant. 
	\begin{figure}
		\centering
		\includegraphics[width=0.8\textwidth]{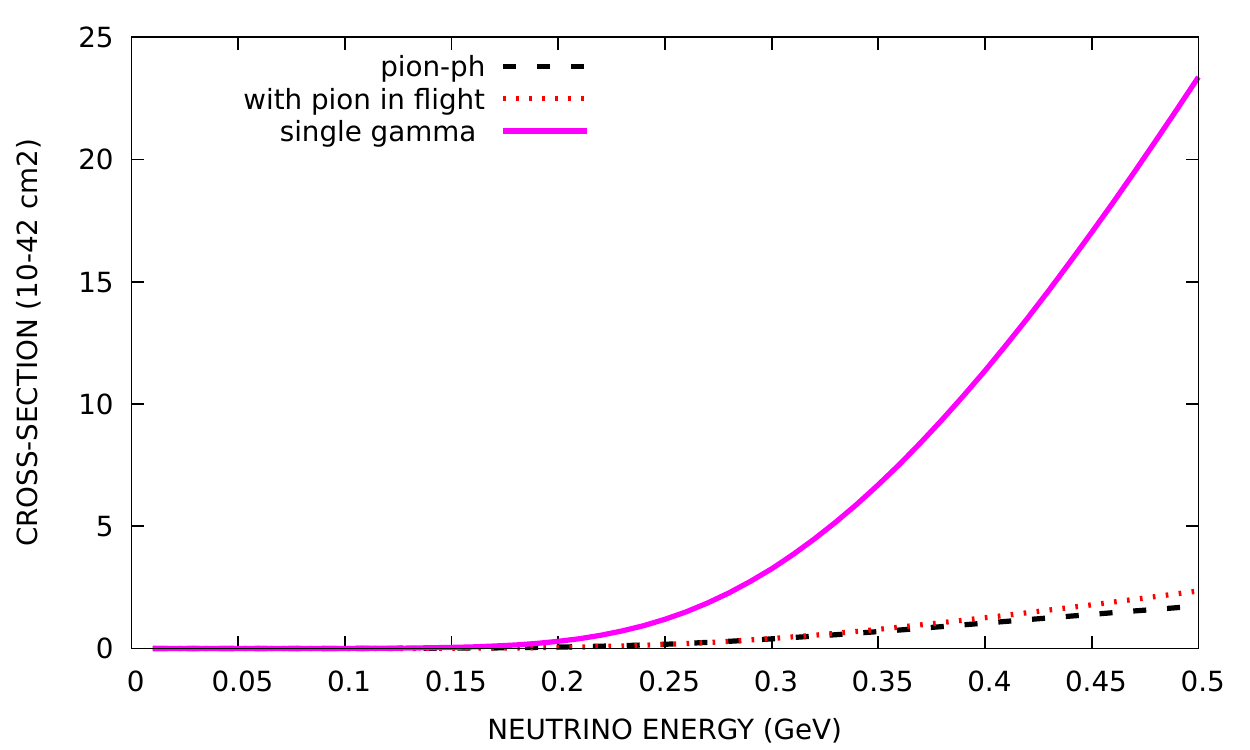}
		\caption{Full line: single  $\gamma$ cross section  for $^{12}C$ (graph of Fig. 1a) in $10^{-42} cm^2$ versus neutrino energy in GeV. Dashed line:  $\gamma$-pion/ph cross section for $^{12}C$ with only the contact term (graph of Fig. 1b). Dotted line:  $\gamma$-pion/ph cross section with inclusion of the pion in flight process  (graphs of Fig. 1b + 1c).  }
		\label{f2}
	\end{figure}
	\begin{figure}
		\centering
		\includegraphics[width=0.8\textwidth]{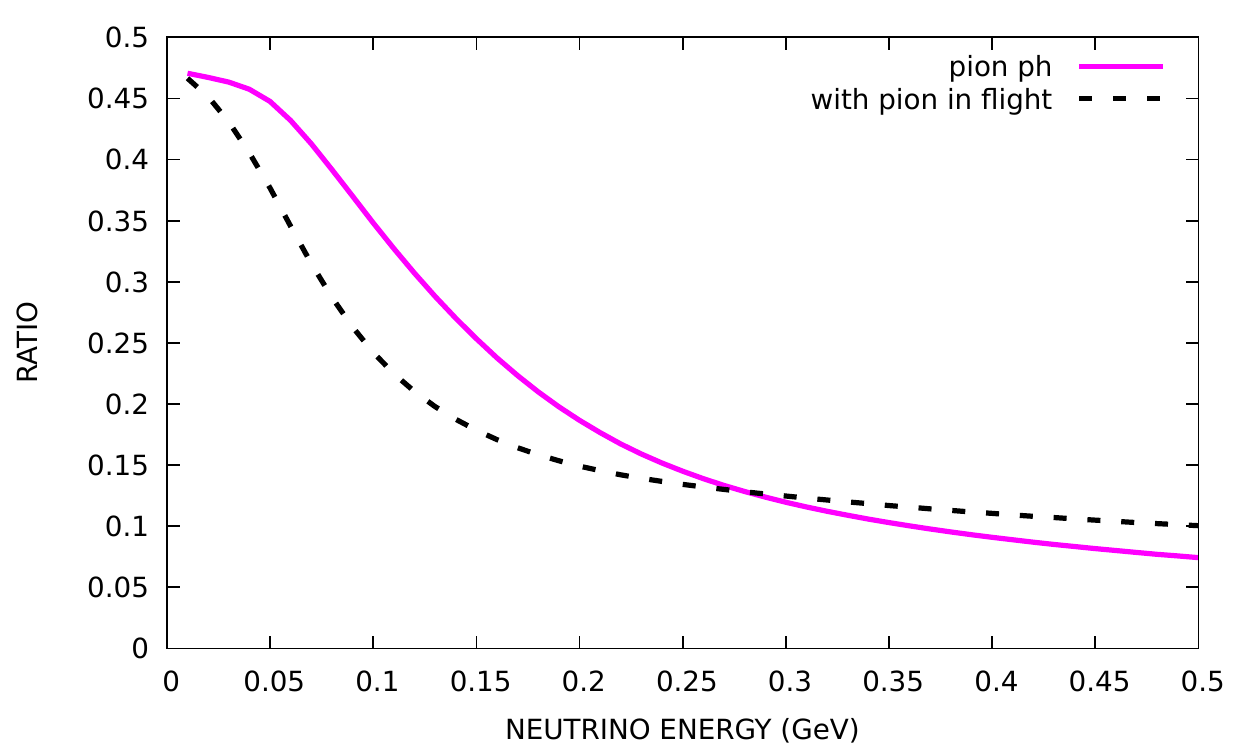}
		\caption{ Ratio of the  $\gamma$-pion/ph and single  $\gamma$ cross sections versus neutrino energy in GeV without (full line) and with (dashed line) the pion in flight process of Fig. 1c. }
		\label{f3}
	\end{figure}
	\begin{figure}
		\centering
		\includegraphics[width=0.8\textwidth]{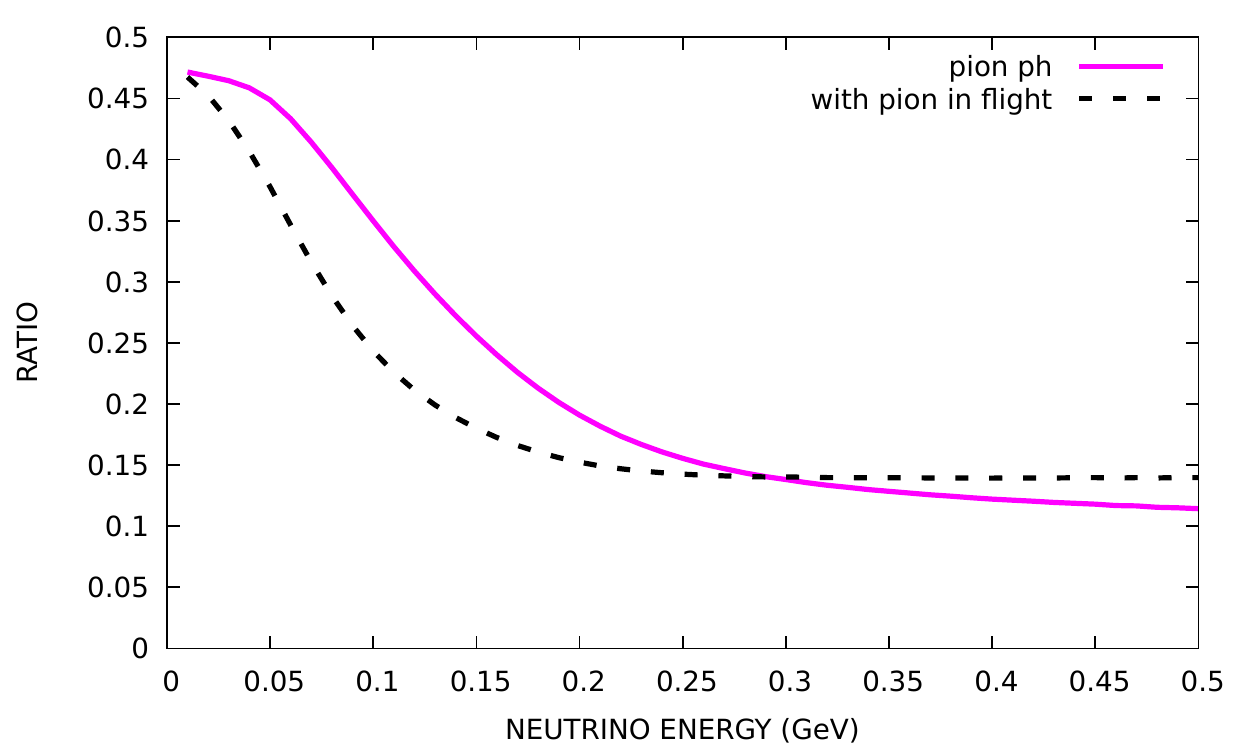}
		\caption{ The same as Fig. 3 but with the inclusion of the 2p-2h contribution in the pion self-energy. }
		\label{f4}
	\end{figure}
	\section{Photon-pion  emission cross-sections  off nuclei}\label{pion cross-section}
	
	As mentioned previously the specific realization of chiral symmetry in the hadronic world implies that for any process involving a vector correlator there is an associated axial correlator process through the emission or the absorption of an s-wave pion; this is referred as the axial-vector correlator mixing effect. This mixing implies that there is   a $N\Delta\gamma\pi$ vertex associated with the direct $N\Delta\gamma$ vertex. There are several ways to derive this vertex. One is to start from an effective chiral theory formulated at the quark level \cite{CE2011} or more directly at the nucleonic level (see Ref. \cite{Holt2016} for a recent review). In both cases the Goldstone pion field $\vec{\phi}(x)$ is introduced through a matrix $U(x)=e^{i\,\vec{\tau}\cdot\vec{\phi}(x)/{f_\pi}}$ having a perfectly well defined transformation law under chiral rotations. Nucleons can be also introduced as heavy sources coupled to pions \cite{BKM95}. The leading term is dictated by chiral symmetry alone :
	\begin{eqnarray}
		{\cal L}^{(1)}_{N}&=&\bar{\psi}_N\left[	i\gamma_{\mu}\left(\partial^{\mu}+ i v^{\mu}\right) + g_A\gamma_{\mu}\gamma^{5} a^{\mu} - M_N\right]\psi_{N}\nonumber\\
		&\simeq&	\bar{\psi}_N\left(	i\gamma_{\mu}\partial^{\mu} - M_N\right)\,-\,
		\frac{g_A}{2 f_\pi} \bar{\psi}_N\gamma_{\mu}\gamma_5\vec{\tau}\psi_N\cdot\partial^{\mu}\vec{\phi}
		-\,\frac{1}{4 f^{2}_{\pi}}
		\bar{\psi}_N \gamma_{\mu}\vec{\tau}\psi_N\cdot\vec{\phi}\times\partial^{\mu}\vec{\phi}\nonumber\\
		\hbox{with} && \xi=\sqrt{U}\qquad v^{\mu}=\frac{i}{2}\left(\xi\partial^{\mu}\xi^\dagger + \xi^\dagger\partial^{\mu}\xi\right)\qquad
		a^{\mu}=\frac{i}{2}\left(\xi\partial^{\mu}\xi^\dagger - \xi^\dagger\partial^{\mu}\xi\right) ,
	\end{eqnarray}
	where the Dirac spinor $\psi_N$ denotes the iso-doublet of nucleons. There are two parameters which are not determined by chiral symmetry : the nucleon mass (in principle in  the chiral limit $M_0$) and  the axial coupling constant, $g_A=1.26$, known from the analysis of neutron beta decay.
	The coupling to the electromagnetic field is simply obtained by gauging the above lagrangian by the appropriate minimal substitution:
	\begin{equation}
		\partial^{\mu}\xi\rightarrow \partial^{\mu}\xi +ie\frac{1+\tau_3}{2} A^\mu\xi\qquad
		\partial^{\mu}\xi^\dagger\rightarrow \partial^{\mu}\xi^\dagger +ie\frac{1+\tau_3}{2} A^\mu\xi^\dagger .
	\end{equation}
	This generates the following Lagrangian:
	\begin{eqnarray}
		{\cal L}_{NN\gamma\pi}&=&-\frac{e}{2}\,g_A\, A_\mu\bar{\psi}_N \,\gamma_{\mu}\gamma^{5}\left( \xi\frac{\tau_3}{2} \xi^\dagger\, -\,\xi^\dagger\frac{\tau_3}{2} \xi\right)\psi_{N}
		\simeq-e\,g_A\, A_\mu\bar{\psi}_N\,\gamma_{\mu}\gamma^{5}\,\frac{\vec\tau}{2f_\pi}\cdot\left(\vec{e}_3\times\vec{\phi}\right)\psi_{N}\nonumber\\
		&=&-i e \frac{g_A}{2f_\pi}A_\mu\,\bar{\psi}_N\,\gamma_{\mu}\gamma^{5}\,\left(\tau^\dagger_{+}\,\phi_+ \,+\,
		\tau^\dagger_{-}\,\phi_-\right)\psi_{N}\nonumber\\
		\hbox{with} &&\tau_{\pm}=\mp\frac{\tau_1\pm i\tau_2}{\sqrt{2}}\qquad \phi_{\pm}=\frac{\phi_1 \pm \phi_2}{\sqrt{2}}\,\,\hbox{creating a}\,\pi^{\pm}.
	\end{eqnarray}
	Hence the matrix element of the electromagnetic interaction between an initial nucleon and and a final state made of a nucleon, a photon and a charged pion $\pi^\alpha$ reads:
	\begin{eqnarray}
		\langle N'(p') ;\gamma({\bf p}_\gamma,\vec\epsilon_\lambda);  \pi^\alpha({\bf p}_\pi)|\int d{\bf x}\,{\bf j}_{em}({\bf x})\cdot{\bf A}({\bf x})\, |N\rangle &=&
		\frac{e}{\sqrt{2p_\gamma V\,2E_\pi V}}\,\frac{g_A}{2f_\pi}\langle p'|\,\int d{\bf x}\,e^{-i({\bf p}_\gamma +{\bf p}_\pi)\cdot{\bf x}}\bar{\psi}_N({\bf x})\,\gamma_{\mu}\gamma^{5}\,\tau^\dagger_{\alpha}{\psi}_N({\bf x})\,|N\rangle\nonumber\\
		&\simeq&\frac{e}{\sqrt{2p_\gamma V\,2E_\pi V}}\,\frac{g_A}{2f_\pi}\,\langle p'|\vec\sigma\cdot\vec\epsilon_\lambda\,\,\tau^\dagger_{\alpha}|N\rangle\,\delta_{{\bf p}_\gamma +{\bf p}_\pi + {\bf p}'-{\bf q}-{\bf p}},
	\end{eqnarray}
	where in the non relativistic approximation of the last line the nucleon states refer to the spin-isospin quantum numbers only. The extension to the $N\Delta\gamma\pi$ vertex is straightforward  with the replacement of the Pauli matrices by the spin and isospin transition operators with the same rescaling of the coupling constants as previously:
	\begin{equation}
		\langle N'(p') ;\gamma({\bf p}_\gamma,\vec\epsilon_\lambda);  \pi^\alpha\left({\bf p}_\pi\right)|\int d{\bf x}\,{\bf j}_{em}({\bf x})\cdot{\bf A}({\bf x})\, |\Delta\rangle 
		=\frac{e}{\sqrt{2p_\gamma V\,2E_\pi V}}\,R_{N\Delta}\,\frac{g_A}{2f_\pi}\,\langle p'|{\bf S}^\dagger\cdot\vec\epsilon_\lambda\,\,T^\dagger_{\alpha}|\Delta\rangle\,\delta_{{\bf p}_\gamma +{\bf p}_\pi + {\bf p}'-{\bf q}-{\bf p}}.\label{contactpion}
	\end{equation}
	After performing again the spin-isospin summation over intermediate delta and final emitted nucleon, we obtain for the photon-pion production cross-section:
	\begin{equation}
		\frac{d\sigma^{\gamma\pi}}{dE'_\nu d\cos\theta}=\left[\frac{G^2_F}{4\pi}\frac{k'}{k}_,8\,kk'\,\left(3\,-\,\cos\theta\right)\right]\,\left[\frac{4}{9}\,R_{N\Delta}^2\,\left(\frac{G_A(Q^2)}{2}\right)^2 
		\right] \left[\frac{1}{2\pi}\,\Gamma^{\gamma\pi}(\omega)\right].\label{crosspion}
	\end{equation}
	The radiative Delta-pion width is:
	\begin{eqnarray}
		\Gamma^{\gamma\pi} (\omega) &=&e^2\,\frac{2}{9}\,R_{N\Delta}^2\, \left(\frac{g_A}{2f_\pi}\right)^2
		\int\frac{d{\bf p}_\gamma}{(2\pi)^3\,2p_\gamma}\int\frac{d{\bf p}_\pi}{(2\pi)^3\,2E_\pi}\,2\pi\,\delta(p_\gamma +E_\pi +\epsilon_{p'}-\omega -\epsilon_p)\nonumber\\
		&\simeq& \frac{e^2}{(2\pi)^3}\,\frac{2}{9}\,R_{N\Delta}^2\, \left(\frac{g_A}{2f_\pi}\right)^2
		\int_0^\infty dp_\gamma\,p_\gamma\,\int_{m_\pi}^\infty dE_\pi\, p_\pi\,\delta(p_\gamma +E_\pi -\omega)\nonumber\\
		&\simeq& \frac{e^2}{(2\pi)^3}\,\frac{2}{9}\,R_{N\Delta}^2\, \left(\frac{g_A}{2f_\pi}\right)^2
		\int_0^\infty dp_\gamma\,p_\gamma\,\int_{m_\pi}^\infty dE_\pi\, p_\pi\,\left(-\frac{2 E_\pi}{\pi}\right)\,
		Im D_{0\pi}\left(\omega - p_\gamma, p_\pi=\sqrt{E^2_\pi-m^2_\pi}\right).\label{gammapi}
	\end{eqnarray}
	In the last two expressions nucleon recoils have been neglected.  The in-medium cross-section is obtained by replacing the bare pion propagator by the in-medium pion propagator according to :
	\begin{equation}
		Im D_{0\pi}(\Omega,p_\pi)=Im \left[\Omega^2-m_\pi^2-p_\pi^2\,+\,i\eta\right]^{-1}\,\,\rightarrow\,\,
		Im D_{\pi}(\Omega,p_\pi)=Im \left[\Omega^2 - m_\pi^2 - p_\pi^2\,- S(\Omega,p_\pi)\right]^{-1},
	\end{equation}
	where $S(\Omega,p_\pi)$ is the irreducible pion-self-energy. This quantity receives contributions from particle-hole  and Delta-hole states corrected by short range screening effects described by three Landau-Migdal parameters \cite{ISW06}: $g'_{NN}=0.7,\, g'_{N\Delta}=0.5,\, g'_{\Delta\Delta}=0.3$, according to:
	\begin{equation}
		S(\Omega,p_\pi)= p^2_\pi\,\tilde{\Pi}_0 (\Omega,p_\pi)
		=p^2_\pi\,	\frac{\Pi_{0N} +\Pi_{0\Delta}+\left(2 g'_{NN}- g'_{N\Delta}-g'_{\Delta\Delta}\right)\,\Pi_{0N}\,\Pi_{0\Delta}}{\left(1- g'_{NN}\Pi_{0N}\right)\left(1- g'_{\Delta\Delta}\Pi_{0\Delta}\right)-\,g'^2_{N\Delta}\Pi_{0N}\,\Pi_{0\Delta}},
	\end{equation}
	with 
	\begin{eqnarray}
		\Pi_{0N} (\Omega,p_\pi )= 4\, \left({g_A\over 2 f_\pi}\right)^2\, 
		v^2(p_\pi)\,
		\int {d{\bf h}\over (2 \pi)^3}& &
		\bigg({\Theta ( p_F-h)\,\Theta (|{\bf h}+{\bf p}_\pi|
			-p_F)\over \Omega\,-\,\epsilon_{{\bf h}+{\bf p}_\pi}\,+\,\,\epsilon_h\,+\,i\eta}
		\nonumber\\
		& &-\,{\Theta (p_F-h)\,\Theta (|{\bf h}-{\bf p}_\pi|
			-p_F)\over \Omega\,+\,\epsilon_{{\bf h}-{\bf p}_\pi}\,-\,\epsilon_h\,-\,i\eta}\bigg),
		\label{PI0N}\end{eqnarray}
	\begin{eqnarray}
		\Pi_{0\Delta} (\Omega, {\bf p}_\pi)= {16\over 9}\,R^2_{N\Delta} \left({g_A
			\over 2f_\pi}\right)^2\, v^2(p_\pi)\,
		\int {d{\bf h}\over (2 \pi)^3}\,\Theta ( p_F-h)& &
		\bigg({1\over \Omega\,-\,\epsilon_{\Delta,{\bf h}+{\bf p}_\pi}\,+\,\epsilon_h\,
			+ i\Gamma_\Delta(\Omega, {\bf h}+{\bf p}_\pi)}
		\nonumber\\
		& &-\, {1\over 
			\Omega\,+\,\epsilon_{\Delta,{\bf h}-{\bf p}_\pi}\,-\,\epsilon_h}\bigg) .
		\label{PI0D}
	\end{eqnarray}
	$v(p_\pi)$ is a dipole $\pi NN$ form factor with cutoff $\Lambda=0.98\,GeV$, as in our previous works \cite{CE2007,MECM2009}. For the imaginary part of the in-medium pion propagator we select only those contributions coming from the dressing of the pion propagator by particle-hole states (see Fig. 1b). Hence the pion  produced is disguised as $p-h$ states which are invisible in the MiniBoone  detector. This simulates a simple gamma process and contributes to an increase of the gamma emission from the Delta decay. The radiative delta-pion width entering the cross section (Eq. (\ref{crosspion})) thus takes the form:
	\begin{equation}
		\Gamma^{\gamma\pi(ph)} (\omega) 
		\simeq \frac{e^2}{(2\pi)^3}\,\frac{2}{9}\,R_{N\Delta}^2\, \left(\frac{g_A}{2f_\pi}\right)^2
		\int_0^\infty dp_\gamma\,p_\gamma\,\int_{m_\pi}^\infty dE_\pi\, p_\pi\,
		\frac{\left(-\frac{2 E_\pi}{\pi}\right)\,p^2_\pi\, Im \tilde{\Pi}_0\left(\omega - p_\gamma, p_\pi=\sqrt{E^2_\pi-m^2_\pi}\right) }
		{\left|\left(\omega-p_\gamma\right)^2 - m^2_\pi -p^2_\pi  -p^2_\pi\, \tilde{\Pi}_0\left(\omega - p_\gamma, p_\pi\right)\right|^2}.
	\end{equation}
	In the relevant energy domain only the ph bubble has an imaginary part: 
	\begin{equation}
		Im\Pi_{0N}(\Omega, {\bf p}_\pi)= -4\,\pi\, \left({g_A\over 2 f_\pi}\right)^2
		\, v^2(p_\pi )\,
		\int {d{\bf h}\over (2 \pi)^3}\,\Theta (p_F-h)\,\Theta (|{\bf h}+{\bf p}_\pi|
		-p_F)\,\delta\left(\Omega\,-\epsilon_{{\bf h}+{\bf p}_\pi}\,+\,\epsilon_h\right).
	\end{equation}\\
	
	The inclusion of the pion in flight term (Fig. 1c) can be achieved by modifying the contact vertex spin operator ${\bf S}^\dagger\cdot\vec\epsilon_\lambda$ appearing in Eq. (\ref{contactpion}) according to \cite{Chanfray1983}:
	\begin{equation}
		{\bf S}^\dagger\cdot\vec\epsilon_\lambda\qquad\rightarrow\qquad{\bf S}^\dagger\cdot\vec\epsilon_\lambda
		-\frac{2\,{\bf S}^\dagger\cdot{\bf t}\, \,{\bf t}\cdot\vec\epsilon_\lambda}{T^2+m_\pi^2},
	\end{equation}
	where ${\bf t}$ and $t_0$ are the momentum and energy ($T^2=t^2-t^2_0$) of the flying pion.  After summation over the photon polarization states and averaging over the photon momentum direction, the net effect is the
	presence in the integrant of $\Gamma^{\gamma\pi} (\omega)$  in Eq. (\ref{gammapi}) of a correction factor $F_{Flight} $ given by:
	\begin{equation}
		F_{Flight}=1 +  \frac{4}{3}\frac{t^2}{T^2+m_\pi^2}\left(\frac{t^2}{T^2+m_\pi^2}-1\right).
	\end{equation}
	If we assume the nucleon recoil momentum to average to zero, we can replace the momentum ${\bf t}$ of the flying pion  by the momentum ${\bf q}$ transferred by the neutrino. As the nuclear response is limited to the low energy domain we can replace the energy  $t_0$ of the flying pion by the photon energy $p_\gamma$. The result of the calculation for the photon-pion emission cross-section without and with the pion in flight terms are shown on Fig. 2. On Fig. 3 we show their ratio with the single gamma cross-section.\\
	
	We have also looked at the  2p-2h absorption mode of the exchanged pion, the dominant mode for physical pions, which would also lead to a simulation of electron rings.   This is achieved by adding to the polarization bubbles $\Pi_{0N,\Delta} (\omega-p_\gamma, {\bf p}_\pi)$ (Eq. \ref{PI0N}) a purely imaginary piece.  According to previous studies \cite{CS93,MECM2009}, we approximate it as linearly growing in the low energy domain, i.e., $\omega-p_\gamma < m_\pi $:
	\begin{equation}
		\Pi_{0 \,2p2h} (\omega-p_\gamma, {\bf p}_\pi)=-i\,4\pi\, Im C_0\,\rho^2 \left(\frac{\omega-p_\gamma}{m_\pi}\right),
	\end{equation}
	with $Im C_0=0.13\, m_\pi^{-6} $ so as to reproduce the absorptive part of the pion-nucleus optical potential. For $\omega-p_\gamma > m_\pi $ we keep it  constant up to a maximum value of $400\,MeV$ for $\omega-p_\gamma > m_\pi $ and zero beyond. 
	\section{Discussion}\label{discussion}
	We show on Fig. 2 the results of the calculation of the cross section for  single gamma emission and for gamma-virtual pion/ph  emission. We also display the result of our calculation when the pion in flight process (Fig. 1c) is introduced on top of the contact term of Fig. 1b. Up to a neutrino energy of about $0.3\, GeV$ the pion in flight contribution is  opposite to that of the contact term. On Fig. 3 and Fig. 4 we display the ratio between the gamma-pion and single gamma cross-sections without and with the 2p-2h absorption mode of the virtual pion. The comparison of these two figures show that the effect of this mode is moderate. \\
	
	In conclusion the  exchange  process that we have introduced is smaller than the single gamma process but nevertheless significant. Here we have considered only the emission from a $N \Delta \gamma $ vertex.  Other similar meson exchange terms may also be at work through more complex pion emission processes. Their complete evaluation  should be done before a definite conclusion can be drawn about the 
	real importance of gamma rings  simulating electron rings production by neutrinos, which could affect some conclusions about the sterile neutrinos. 
	Finally we remind that our evaluation is  restricted to the influence of the pion exchange effects on the average  of the neutrino and antineutrino cross-sections. The investigation on the influence on their difference, which is relevant for CP violation, will be the object of a future work.

	{\bf \Large{}}
	\appendix 
	
	\section{Details for the calculation of the single gamma cross-section}
	The calculation of the single gamma cross-section involves the axial vector matrix element (Eq. \ref{axialmatrix}) and the electromagnetic matrix element (Eq. \ref{elec}). When re-injected in the expression of the cross-section (Eq. \ref{sigmagen}), this requires a summation over the spin-isospin states of the intermediate Delta and of the final emitted nucleon according to :
	\begin{eqnarray}
		&&\sum_{M'_S,M'_T}\sum_{M_S,M_T}\sum_\lambda\sum_{j,k}\,\langle p : m_s, m_t|S_j^\dagger\,T_3^\dagger|M'_S,M'_T\rangle \int\frac{d{\hat p}_\gamma}{4\pi} \sum_{m'_s,m'_t,\lambda}
		\langle M'_S,M'_T| \left({\bf S}\times{\bf p}_\gamma\right)\cdot\vec\epsilon^*_\lambda\,T_3\,|m'_s,m'_t\rangle\nonumber\\
		&&\langle m'_s,m'_t|\left({\bf S}^\dagger\times{\bf p}_\gamma\right)\cdot\vec\epsilon_\lambda\,T_3^\dagger\,|M_S,M_T\rangle\,\,\,\,\langle M_S,M_T|S_k\,T_3|p : m_s, m_t\rangle\,L^{jk}\nonumber\\
		&&=\sum_{M'_S,M'_T}\sum_{M_S,M_T}\sum_{j,k}\,\langle p : m_s, m_t|S_j^\dagger\,T_3^\dagger|M'_S,M'_T
		\rangle \,\,\,\int\frac{d{\hat p}_\gamma}{4\pi}\,p^2_\gamma\,\frac{4}{9}\,\delta_{M_S,M'_S}
		\,\delta_{M_T,M'_T}\,\,\,\langle M'_S,M'_T|S_k\,T_3|p : m_s, m_t\rangle\,L^{jk}\nonumber\\
		&&=\int\frac{d{\hat p}_\gamma}{4\pi}\,\,\frac{4p^2_\gamma}{9}\,\,
		\langle p : m_s, m_t|\left(\frac{2}{3}\delta_{jk}-\frac{i}{3}\varepsilon{jkl}\,\sigma_l\right)\frac{2}{3}|p : m_s, m_t\rangle\,L^{jk} = \int\frac{d{\hat p}_\gamma}{4\pi}\,\,\frac{4p^2_\gamma}{9}\,\frac{4}{9}\, (L^{11}+L^{22}+L^{33})\nonumber \\
		&&=\int\frac{d{\hat p}_\gamma}{4\pi}\,\,\frac{4p^2_\gamma}{9}\frac{4}{9}\,
		\left[8\,kk'\,\left(3\,-\,\cos\theta\right)\right].
	\end{eqnarray}
	Grouping  all the terms together, the cross section per nucleon for gamma production given in Eq. \ref{sigmasingle} follows.


\begin{thebibliography}{99}
		\bibitem{MiniBooNe2009} A. Aguilar-Arevalo, et al., MiniBoone Collaboration, Phys. Rev. Lett. 102, 101802 (2009).
		\bibitem{MiniBooNe2013} A. Aguilar-Arevalo, et al., MiniBoone Collaboration, Phys. Rev. Lett. 110, 161801 (2013).
		\bibitem{MiniBooNe2018} A. Aguilar-Arevalo, et al., MiniBoone Collaboration, Phys. Rev. Lett. 121, 221801 (2018).
		\bibitem{Hill2010} R.J. Hill, Phys. Rev.D 81, 013008 (2010).
		\bibitem{Hill2011} R.J. Hill, Phys. Rev.D 84, 017501 (2011).
		\bibitem{Serot2012} B.D. Serot and X. Zhang, Phys. Rev. C 86, 015501 (2012).
		\bibitem{Serot2013} X. Zhang and B.D. Serot, Phys. Lett. B, 719 (2013).
		\bibitem{Wang2014} E. Wang, L. Alvarez-Ruso and J. Nieves, Phys. Rev. C 89, 015503 (2014).
		\bibitem{Wang2015} E. Wang, L. Alvarez-Ruso and J. Nieves, Phys. Lett. B 740, 16 (2015).
		\bibitem{Katori2020} T. Katori for the MiniBoone collaboration, arXiv:hep-ex/2010.06015. 
		\bibitem{CDER99} G. Chanfray, J. Delorme, M. Ericson and M. Rosa-Clot, Phys. Lett. B 455, 39 (1999). 
		\bibitem{DEI90}  M. Dey, V.L. Eletsky and B.L. Ioffe, Phys. Lett. B 252, 620 (1990).
		\bibitem{CS93} G. Chanfray and P. Schuck, Nucl. Phys. A 555, 329 (1993).   
		\bibitem{CRW96}G. Chanfray, R. Rapp and J. Wambach, Phys. Rev. Lett. 76, 368 (1996). R. Rapp, G. Chanfray and J. Wambach, Nucl. Phys. A 617, 472 (1997).
		\bibitem{Oset82} E. Oset, H. Toki and W. Weise, Phys. Rept. 83, 281 (1982).
		\bibitem{CE2011} G. Chanfray and M. Ericson, Phys. Rev C 75, 015204 (2011).		
		\bibitem{Holt2016} W. Holt, M. Rho and W. Weise, Phys. Rept. 621, 2 (2016).	
		\bibitem{BKM95} V. Bernard, N. Kaiser and U.-G. Meissner, Int. J. Mod. Phys. E 4,  193 (1995).	
		\bibitem{ISW06} M. Ichimura, H. Sakai and T. Wakasa, Prog. Part. Nucl. Phys. 56, 446 (2006).		
		\bibitem{CE2007}G. Chanfray and M. Ericson, Phys.Rev. C75  015206 (2007).
		\bibitem{MECM2009} M. Martini, M. Ericson, G. Chanfray and J. Marteau,  Phys.Rev. C80 065501 (2009).
		\bibitem{Chanfray1983} G. Chanfray and J. Delorme, Phys. Lett. B 129, 167 (1983);  G. Chanfray, Nucl. Phys. A 429, 389 (1984). 
		
		
	\end{thebibliography}
\end{document}